\documentclass[pra,twocolumn,showpacs,preprintnumbers,amsmath,amssymb,superscriptaddress]{revtex4}

\usepackage{graphicx}
\usepackage{dcolumn}
\usepackage{bm}
\usepackage[dvips]{epsfig}

\begin{document}



\title{Noisy channel effect on quantum correlations of two relativistic particles}

\author{M.~Mahdian}
\altaffiliation{%
Author to whom correspondence should be addressed; electronic
mail:  mahdian@tabrizu.ac.ir}

\affiliation{%
Faculty of Physics, Theoretical and astrophysics department , University of Tabriz, 51665-163 Tabriz, Iran}

\author{T.~Makaremi}

\affiliation{%
Department of Physics, Faculty of Science, University of Kurdistan,  Sanandaj, Iran}

\author{Sh.~Salimi}

\affiliation{%
Department of Physics, Faculty of Science, University of Kurdistan,  Sanandaj, Iran}
%



\begin{abstract}

 We study the quantum correlation dynamics of two relativistic particles which is transmitted through one of the Pauli
 channels $ \sigma_{x},  \sigma_{y},$ and  $\sigma_{z}$.
 We compare sudden death and robustness of entanglement and geometric discord and quantum
 discord of two relativistic particles under noisy Pauli channels.
 we find out geometric discord and quantum discord may be more robust than entanglement against
 decoherence.
\end{abstract}
%




\maketitle

\section{ Introduction}

In quantum information theory, quantifying of quantum correlation
plays a important role. However, despite of researcher's
attention that have more
focused on entanglement, the measures of non-classical
correlation in quantum information theory is not restricted on
quantum entanglement. Recently, quantum discord is
introduced as a quantum correlation that isn't vanished for
mixed separable state, against quantum entanglement.
Quantum discord define as difference of two natural quantum
expansion of classical
mutual information \cite{GAlber}-\cite{DGirolami}.
Because of hard work to obtain quantum discord and most work has
been done on specific models such as x-state\cite{Amazaher,DGirolami,SLuo},
Dakic, Vedral and Brukner \cite{DVB}introduced another measure so-called
geometric discord for a bipartite state. Next, Girolami and Adesso introduced function of
geometric discord as an upper bound for quantum discord for two qubit states.
Moreover, the quantifying of quantum correlations have been
interested in some physical models and information processing such
as relativity particles \cite{jafarizadehMahdian1}-
\cite{Doyeol}.
These efforts offer new prospects for quantum information researchers.
For example, they find out environments as sources of decoherence act on
open quantum systems. Therefore the study of quantum entanglement
and specially quantum discord as a  universal quantum resource has
more and more importance. Based on it,
researchers have investigated different quantum noises such as Pauli
channels effects on quantum correlations of
quantum states such as Greenber-Horne-Zeilinger (GHZ) state and
Werner state,etc \cite{M.Siomau, I.CP.AA.k.P}.
In order to provide a precise definition we can say quantum
channels are completely positive and
trace preserving maps between spaces of operators
\cite{Mahdian.yousefjani.salimi}.
The investigation of this case include Markovian and non-Markovian
channels depend on the
interaction of environment and system. On one hand Markovian channels
describes memoryless environment that is obtained by solving a master equation for the
reduced
density matrix with Lindblad structure.
On the other hand environment has memory in non-Markovian channels therefore
the master equations for this case is complicated integro-differential and rarely exactly solvable
\cite{T.YuJ.H.E, T.Yu.J.H.Eberlyop, A.K.R}.

In this paper we compare the quantum entanglement with geometric discord and quantum discord
dynamics of two relativistic particles under Pauli noisy channels
. We find out that under Pauli channels $ \sigma_{x}, \sigma_{y}, \sigma_{z} $ quantum entanglement
and geometric discord and quantum discord tend to death asymptotically related on the velocity of observer.
This behavior occur with more gradient for the lower speed observers in
$ 0 < \theta< \frac{\pi}{2} $ and in $ \frac{\pi}{2} < \theta< \pi $ these are vice versa.
The quantum entanglement sudden death(ESD) occur $ \sigma_{x}, \sigma_{z} $ and we survey the time of ESD for them.

This paper is organized as follows: In Sec. 2, we review some
of basic definitions to
introduction to concurrence as a measure of entanglement
and geometric discord and quantum discord
. In Sec. 3 we discuss
the computation of entanglement (by
using the concurrence) and quantum discord
two maximally spin entangled states, respectively. In Sec. 4, we
investigate how two relativistic particles as initial state evolute under noisy channels. In Sec. 5 , 6, 7 we investigate
noisy channel effect on entanglement and geometric discord and quantum
discord of two relativistic particles state.
Conclusion are then presented in Sec. 6.

\section{ Basic definition}
\subsection{ Concurrence}
In this subsection, we present the concurrence which is a good
measure of entanglement for two qubit subsystems. The spin-flipped
state for two qubits is
\begin{equation}\label{ro tilda}
\tilde \rho  = (\sigma_{y}\otimes\sigma_{y})\rho^{\dagger}(\sigma_{y}\otimes\sigma_{y}).
\end{equation}
Note that the complex conjugate is taken in the computational basis
$ {|0 0\rangle, |0 1\rangle, |1 0\rangle, |1 1\rangle} $. The
concurrence is defined as follows
\begin{equation}\label{concurrence}
C= max \Big[0 , (\chi_{1}-\chi_{2}-\chi_{3}-\chi_{4})\Big] ,
\end{equation}
where the $ \chi_{i}$ are the square roots of the eigenvalues of
the matrix $ R  = \rho \tilde{\rho} $ \cite{31,32}.
\subsection{Geometric Discord}

Geometric discord is identified as the closest distance between an
arbitrary state and a classical-quantum state (zero discord)\cite{DGirolami,DVB}.
Therefore, the density matrix in terms of Bloch states are expanded as follows:
\begin{equation}\label{bloch matrix}
    \rho=\frac{1}{4}(I\otimes I+\sum_{i=1}^{3}x_{i}\sigma_{i}\otimes I+
    \sum_{j=1}^{3}y_{i}I\otimes \sigma_{i}+\sum_{i,j=1}^{3}t_{ij}\sigma_{i}\otimes \sigma_{j})
\end{equation}
where $ x_{i} $ and $ y_{i} $ are the three-dimensional Bloch vectors
of subsystems and $ t_{ij} $ is the elements of correlation matrix T.
Therefore we can obtain geometric discord as:
\begin{equation}\label{DG}
    DG(\rho)=\frac{1}{4}(\|\overrightarrow{y}\overrightarrow{y}^{T}\|_{2}+\|T\|_{2}^{2}-k)
\end{equation}
where k being the largest eigenvalue of matrix  $ \overrightarrow{y}\overrightarrow{y}^{T}+T^{T}T $
 and $ \|.\|_{2}^{2} $ is the norm of Hilbert-Schmidt.
\subsection{Quantum Discord}

Quantum discord is defined in terms of difference
between two expressions for
mutual information that are same in classical form
but are difference in quantum form.
These two classical expressions are:
\begin{equation}\label{Iclassic}
    I(A,B)=H(A)+H(B)-H(A,B),
\end{equation}
\begin{equation}\label{Jclassic}
    J(A,B)=H(A)-H(A|B).
\end{equation}
Where H(.) is Shannon entropy. We can see these two
classical equation in above are equivalent.
In quantum case, we are used an quantum entropy are so
called Von Neumann entropy $ S(\rho)$ .
Therefore we can define two expressions for quantum mutual information:
\begin{equation}\label{Iquantum}
    I(\rho)= S(\rho^{A})+S(\rho^{B})-S(\rho),
\end{equation}
\begin{equation}\label{Jclassic}
    J_{A}(\rho)= S(\rho^{B})-S(\rho^{B}|\rho^{A})
\end{equation}
Despite of classical case, two equations in above are inequivalent.
The difference between the two expressions $ I(\rho)-J_{A}(\rho) $
defines the basis-dependent quantum discord which is asymmetrical
in the sense that $  D_{A}(\rho) $ can differ from $ D_{B}(\rho) $
\cite{SLuo}. Classical correlation is defined as:
\begin{equation}\label{classical correlation}
     CC(\rho)= \max_{{\{\pi_{j}^{A}}\}} J_{{\{\pi_{j}^{A}\}}}(\rho).
\end{equation}
Where maximum is over the set of all possible projective measurements
on to the eigen basis \cite{nanluo}. According to
Refs.\cite{HOllivierZurek, LHendersonVedral, DGirolami},
we can define quantum discord as:
$$
    D_{A}=I(\rho)- max_{{{\{\pi_{j}}^{A}\}}} J_{{\{\pi_{j}}^{A}\}}(\rho)=S(\rho_{A})-S(\rho)
    $$
    \begin{equation}\label{quantum discord definition}
      +min_{{\{\pi_{j}^{A}\}}}S(\rho_{B|{\{\pi_{j}^{A}\}}}).
\end{equation}
\section{Two relativistic  particles}

Let us consider a maximally spin- entangled state for two $ s=1/2 $
fermions or two photons A and B . Consider two particles are far
apart
\begin{equation}\label{bell state}
 |\psi_{p}^{-}\rangle:=\frac{1}{\sqrt{2}}(\psi_{\uparrow}^{a} (p_{a})
 \psi_{\downarrow^{b}}(p_{b})-
  \psi_{\downarrow}^{a}(p_{a}) \psi_{\uparrow}^{b} (p_{b})),
\end{equation}
where $ p_{a} $ , $ p_{b} $ are the corresponding momentum vectors
of particles A, B and $\varphi_{\uparrow}(p)$ ,
$\varphi_{\downarrow}(p)$ are spins of its particles which are in
direction of z-axis. We choose equal interaction angles for the two
particles, $ \alpha_{p_{i}}^{a}=\alpha_{p_{i}}^{b} $ , as a natural
simplification.
 Based on computation in Ref.\cite{16}, we  obtain the
 density matrix for this x-state as
 \begin{equation}\label{ro bell state}
 \rho(0)=\left(
        \begin{array}{cccc}
          \eta & 0 & 0 & \eta\\
          0 & \frac{1}{2}-\eta & \eta-\frac{1}{2} & 0 \\
          0 & \eta-\frac{1}{2} &  \frac{1}{2}-\eta & 0 \\
          \eta & 0 & 0 & \eta \\
        \end{array}
      \right),
 \end{equation}
 where $ \theta=(\alpha_{1}-\alpha_{2})$,
  and
 $ \eta=\frac{1}{4}\sin^{2}\theta $.
 It is easy to obtain the concurrence and geometric
 discord and quantum discord of this system
 as follow:
 $$
 C= 2(|\xi|-|\eta|),
 $$
 $$
 DG=\frac{1}{2}(1-4\eta)^{2},
 $$
 \begin{equation}\label{before noise}
 D= -1+\frac{8(\xi\ln8\xi+\eta\ln8\eta)}{\ln16}
 \end{equation}
 respectively, where $ \xi =\frac{1}{8}(3+\cos2\theta) $.

\begin{figure}[htb]
\begin{center}
\includegraphics{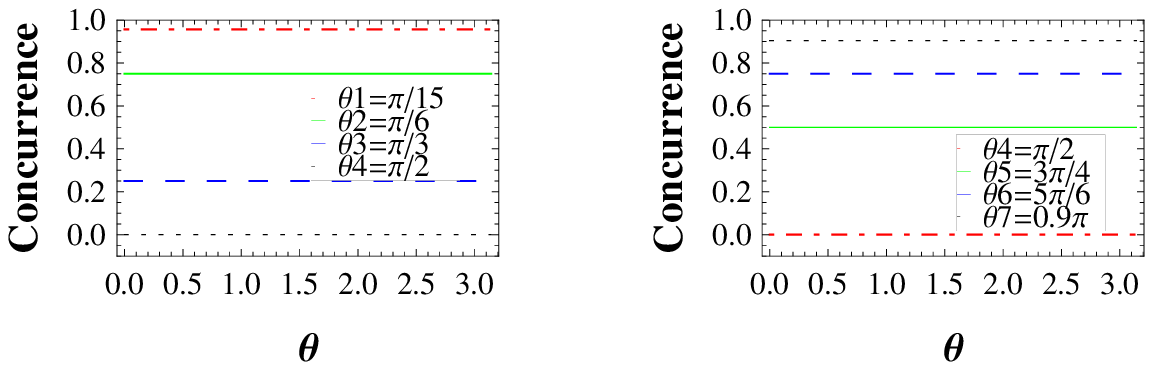}
      \vspace{6cm}
\caption[]{Concurrence for $ 0 <\theta <\frac{\pi}{2}$
(left panel) and concurrence for $ \frac{\pi}{2}<\theta <\pi $
(right panel) measures versus $ \theta $ for different values of $ \theta $
before noise is sketched in this figure. As we seen by increase of $ \theta $
 in $ 0 <\theta <\frac{\pi}{2}$ concurrence is decreased but for
  $ \frac{\pi}{2}<\theta <\pi $ is vice versa.}
\label{fig1}
\end{center}
\end{figure}

\begin{figure}[htb]
\begin{center}
\includegraphics{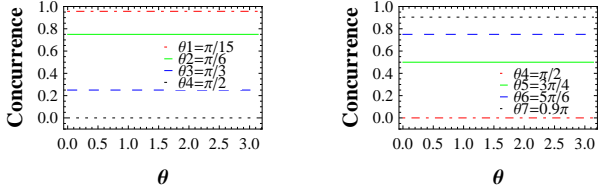}
      \vspace{6cm}
\caption[]{Geometric discord (DG) for $ 0 <\theta <\frac{\pi}{2}$
(left panel) and geometric discord for $ \frac{\pi}{2}<\theta <\pi $
(right panel) measures versus $ \theta $ for different values of $ \theta $
before noise is sketched in this figure. As we seen by increase of $ \theta $
in $ 0 <\theta <\frac{\pi}{2}$ geometric discord is decreased but for
$ \frac{\pi}{2}<\theta <\pi $ is vice versa.}
\label{fig2}
\end{center}
\end{figure}

\begin{figure}[htb]
\begin{center}
\includegraphics{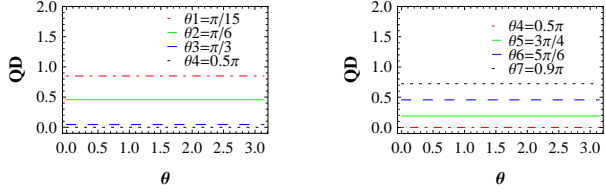}
      \vspace{6cm}
\caption{Quantum discord(QD) for $ 0 <\theta <\frac{\pi}{2}$ (left panel)
and quantum discord for $ \frac{\pi}{2}<\theta <\pi $ (right panel) measures versus $ \theta $
 for different values of $ \theta $ before noise is sketched in this figure. As we seen
  by increase of $ \theta $ in $ 0 <\theta <\frac{\pi}{2}$ quantum discord is
  decreased but for $ \frac{\pi}{2}<\theta <\pi $ is vice versa.}
\label{fig3}
\end{center}
\end{figure}

 Dependence of all of these three measures of quantum correlation are
 sketched in Figs. \ref{fig1}- \ref{fig3}.
 According to these figures  we can find out geometric discord and quantum
 discord of this case depend
 on angles of between momentum and spin of each particles
 as well as quantum entanglement. We can see in these figures
 that quantum entanglement and geometric discord and quantum
 discord in interval $ 0<\theta<\frac{\pi}{2} $ are decreased by increase
 $\theta$ and for interval $ \frac{\pi}{2}<\theta<\pi $ are vice versa.

 \section{Time evolution of two relativistic particle
 state transmitted through noisy channel}
 For open systems that are coupled to their environment, decoherence occurs.
 For this purpose, researchers has investigated a lot of noise models such
 as Pauli channels, in recent decades.
 In this Section,
 we  shall give a survey of the various types of channels which can be effected on
 the quantum dynamics of open system as an initially entangled state $ \rho(0) $
 for two relativistic particles
 that was supposed to be transmitted through these channels for the time t,
 respectively and it's time evolution $ \rho(t) $ will obtain as a
 solution of a master equation:
 \begin{equation}\label{master equation}
    \frac{\partial\rho}{\partial t}=-i[H_{s},\rho]+
    \sum_{k}\frac{1}{2}\gamma_{k}(2L_{k}\rho L_{k}^{\dag}-
    \{L_{k}^{\dag}L_{k},\rho\}).
 \end{equation}
 In this equation $ L_{k}\equiv 1 \otimes\sigma_{k} $ describes the
 decoherence of first qubit under $ \sigma_{k} $ and $ \gamma_{k}$ is the coupling constant
 and
 this master equation should be solved at $ t>0 $ \cite{M.Siomau}. In this paper we
assume $ H_{s}=0 $ and one of particles is subject to the local
noises\cite{Yeo}.

\section{Time evolution of quantum entanglement of two relativistic particles
 state transmitted under Pauli channels}
\subsection{Pauli channel $ \sigma _{x} $}

\begin{figure}[htb]
\begin{center}
\includegraphics{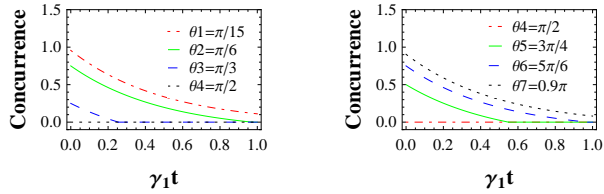}
      \vspace{6cm}
\caption{Concurrence for $ 0 <\theta <\frac{\pi}{2}$ (left panel)
and concurrence for $ \frac{\pi}{2}<\theta <\pi $ (right panel) measures
versus $\gamma_{1}t$ for different values of $ \theta $ under Pauli noisy channel
$ \sigma_{x} $ is sketched in this figure. As we seen
in $ 0 <\theta <\frac{\pi}{2}$ ESD is more sudden for the lower speed observer
 but for $ \frac{\pi}{2}<\theta <\pi $ is vice versa.}
\label{fig4}
\end{center}
\end{figure}

If one of two relativistic particles with initial state as
Eq.(\ref{ro bell state}) is subject to a local bit-flip noise,
the time evolution is obtain by the solution of the master equation
Eq.(\ref{master equation}) with Lindblad operator $ L_{k}\equiv 1 \otimes\sigma_{x} $.
So after transmission of two relativistic particles through the Pauli
channel $ \sigma_{x} $ the density matrix is given as follows:
\begin{equation}\label{density matrix after sigma x}
   \rho_{x} =\frac{1}{4}\left(
       \begin{array}{cccc}
         1-\lambda & 0 & 0 & 1+\mu_{1}-4\xi \\
         0 & 1+\lambda & 1-\mu_{1}-4\xi & 0 \\
         0 & 1-\mu_{1}-4\xi & 1+\lambda & 0 \\
         1+\mu_{1}-4\xi & 0 & 0 & 1-\lambda \\
       \end{array}
     \right)
\end{equation}
where $ \rho_{x} $ is the density matrix after transmission through Pauli
channel $ \sigma_{x} $ and $ \mu_{1}= \exp(-2\gamma_{1}t) $ and $ \lambda_{1}= \mu_{1}(1-4\eta) $.
 We consider concurrence as a measure of quantum entanglement in this section.
 Therefore, for this case concurrence is given as:
 \begin{equation}\label{concurrence x}
    C = Max\left[0, \frac{1}{2}[\mu_{1}+\lambda_{1}+4\left(8\xi^{2}-3\xi+1\right)]\right].
 \end{equation}
As we see in Figs. \ref{fig4} crossing the Pauli channel
$ \sigma_{x} $ concurrence for the
observer in the interval $ 0 < \theta< \frac{\pi}{2} $
is decreased by increase of velocity of observer. It means that in this
interval ESD is happened for higher speed observers more sudden.
On the other hand, in the interval $ \frac{\pi}{2} < \theta< \pi  $
concurrence is increased by increase of velocity of observer. In the other
words in this interval the concurrence reduction with observer speed is slower.
It means that ESD occur for lower speed observers faster.
We can obtain the ESD time for this case as:
\begin{equation}\label{esd time}
    t_{ESD}=\frac{1}{\gamma_{1}}\ln(\sqrt{\frac{\xi}{\eta}})
\end{equation}
\subsection{Pauli channel $ \sigma_{y} $}

\begin{figure}[htb]
\begin{center}
\includegraphics{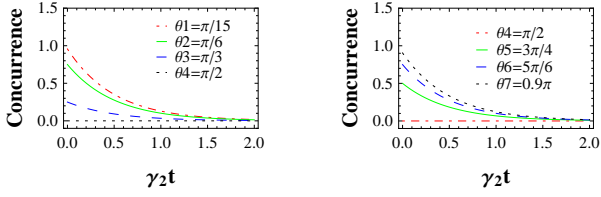}
      \vspace{6cm}
\caption{Concurrence for $ 0 <\theta <\frac{\pi}{2}$ (left panel)
and concurrence for $ \frac{\pi}{2}<\theta <\pi $ (right panel) measures
versus $\gamma_{2}t$  for different values of $ \theta $ under Pauli noisy channel
$ \sigma_{y} $ is sketched in this figure. As we seen ESD don't occur but all of them
tend to quantum entanglement death asymptotically
  for lower speed observers more
sudden in $ 0 <\theta <\frac{\pi}{2}$ but for $ \frac{\pi}{2}<\theta <\pi $ is vice versa.}
\label{fig5}
\end{center}
\end{figure}

If one of two relativistic particles with initial state as
Eq.(\ref{ro bell state}) is subject to a local Pauli channel $ \sigma_{y} $,
the time evolution is obtain by the solution of the master equation
Eq.(\ref{master equation}) with Lindblad operator $ L_{k}\equiv 1 \otimes\sigma_{y} $.
So after transmission of two relativistic particles through the Pauli
channel $ \sigma_{y} $ the density matrix is given as follows:
\begin{equation}\label{density matrix afer Y}
    \rho_{y}= \frac{1}{4}\left(
                \begin{array}{cccc}
                  1-\lambda_{2} & 0 & 0 & 1-\lambda_{2} \\
                  0 & 1+\lambda_{2} & -(1+\lambda_{2}) & 0 \\
                  0 & -(1-\lambda_{2}) & 1-\lambda_{2} & 0 \\
                  1-\lambda_{2} & 0 & 0 & 1-\lambda_{2} \\
                \end{array}
              \right)
\end{equation}
where $ \rho_{y} $ is the density matrix after transmission through Pauli
channel $ \sigma_{y} $ and $ \mu_{2}= \exp(-2\gamma_{2}t) $ and $ \lambda_{2}= \mu_{2}(1-4\eta) $.
So, for this case we can obtain concurrence as:
 \begin{equation}\label{concurrence y}
    C = \frac{1}{2}\left(\left|\lambda_{2}+1\right|-\left|\lambda_{2}-1\right|\right).
 \end{equation}
As we see in Figs. \ref{fig5} under the Pauli channel
$ \sigma_{y} $ the gradient reduction of concurrence for the
observer in the interval $ 0 < \theta< \frac{\pi}{2} $
is more for lower speed observers. It means that in this
interval system tend to ESD  for lower speed observers more sudden.
On the other hand, in the interval $ \frac{\pi}{2} < \theta< \pi  $
the concurrence gradient reduction is more for higher speed observers. In the other
words if the observer tend faster to ESD occur for him faster in this interval.

\subsection{Pauli channel $ \sigma_{z} $}

\begin{figure}[htb]
\begin{center}
\includegraphics{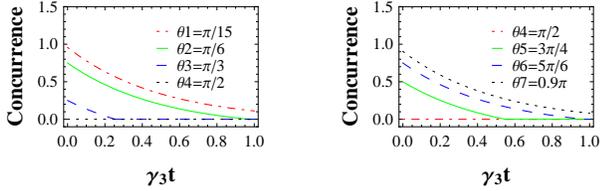}
      \vspace{6cm}
\caption{Concurrence for $ 0 <\theta <\frac{\pi}{2}$ (left panel)
and concurrence for $ \frac{\pi}{2}<\theta <\pi $ (right panel) measures
versus $\gamma_{3} t $  for different values of $ \theta $ under Pauli noisy channel
$ \sigma_{z} $ is sketched in this figure. As we seen ESD
in $ 0 <\theta <\frac{\pi}{2}$ is happen for lower speed observers more
sudden but for $ \frac{\pi}{2}<\theta <\pi $ is vice versa.}
\label{fig6}
\end{center}
\end{figure}

If one of two relativistic particles with initial state as
Eq.(\ref{ro bell state}) is subject to a local dephasing noisy channel,
the time evolution is obtain by the solution of the master equation
Eq.(\ref{master equation}) with Lindblad operator $ L_{k}\equiv 1 \otimes\sigma_{z} $.
So after transmission of two relativistic particles through the Pauli
channel $ \sigma_{z} $ the density matrix is given as follows:
\begin{equation}\label{ro z}
    \rho_{z}=\left(
               \begin{array}{cccc}
                 \eta & 0 & 0 & \eta\mu_{3} \\
                 0 & \xi & -\xi\mu_{3} & 0 \\
                 0 & -\xi\mu_{3} & \xi & 0 \\
                 \eta\mu_{3} & 0 & 0 & \eta \\
               \end{array}
             \right)
\end{equation}
where $ \rho_{z} $ is the density matrix after transmission through dephasing
channel and $ \mu_{3}= \exp(-2\gamma_{3}t) $ and $ \lambda_{3}= \mu_{3}(1-4\eta) $.
According to density matrix in above concurrence could be obtain as:
 \begin{equation}\label{concurrence z}
    C = Max\left[0,2\left(\left|\mu_{3}\xi\right|-\eta\right)\right].
 \end{equation}
According to Fig.\ref{fig6} we can find out in the interval
$ 0<\theta<\frac{\pi}{2} $ the concurrence leads to ESD
faster by increase speed of observer but for the observers in the interval
$ \frac{\pi}{2}<\theta<\pi $ ESD is happen later for high speed observer.
The time of ESD could be obtain as:
\begin{equation}\label{time deathz}
    t_{ESD}=\frac{1}{4\gamma_{3}}\ln\left(1-2\csc^{2}\theta\right)^{2}.
\end{equation}

\section{Time evolution of geometric discord of two relativistic particle
 state transmitted under Pauli channels}
\subsection{Pauli channel $ \sigma _{x} $ , $ \sigma _{z} $}

\begin{figure}[htb]
\begin{center}
\includegraphics{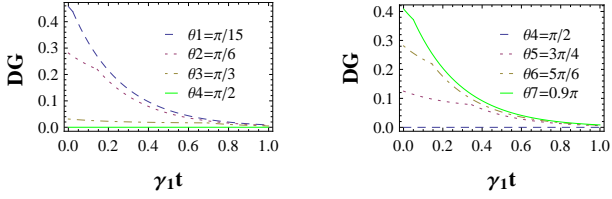}
      \vspace{6cm}
\caption{Geometric discord(DG) for $ 0 <\theta <\frac{\pi}{2}$ (left panel)
and geometric discord for $ \frac{\pi}{2}<\theta <\pi $ (right panel) measures
versus $\gamma_{1(3)} t $  for different values of $ \theta $ under Pauli noisy channel
$ \sigma_{x(z)} $ is sketched in this figure. As we seen geometric discord sudden death
don't occur but geometric discord tend to death in for
lower speed observers more sudden $ 0 <\theta <\frac{\pi}{2}$
but for $ \frac{\pi}{2}<\theta <\pi $ is vice versa.}
\label{fig7}
\end{center}
\end{figure}

If the time evolution of density matrix when one of two
relativistic particles is subject to a local bit-flip noise and dephasing channel
are in the form of \ref{density matrix after sigma x}, \ref{ro z}, respectively then
the geometric discord is obtained as:
$$
    DG_{x(z)} = \frac{1}{4}\left[(1-4\eta)^{2}+\mu_{1(3)}^{2}(1+(1-4\eta)^{2})\right]
    $$
\begin{equation}\label{DGx}
    - \frac{1}{4}Max\left[\mu_{1(3)}^{2}, (1-4\eta)^{2}, \mu_{1(3)}^{2}(1-4\eta)^{2}\right]
\end{equation}
where $ \mu_{1(3)} $ is related on when
Pauli channels  $ \sigma _{x(z)} $ act on system.
According to Fig.\ref{fig7} we could see in the interval
$ 0<\theta<\frac{\pi}{2}$  geometric discord
for the observers with a lower rate is more.
So geometric discord death in this interval
occur more sudden for the lower speed observers
but in the interval $ \frac{\pi}{2}<\theta<\pi $
geometric discord death occur more sudden for higher
speed observers.

\subsection{Pauli channel $ \sigma _{y} $}

If the time evolution of density matrix when one of two
relativistic particles is subject to a Pauli channel  $ \sigma _{y} $,
is in the form of \ref{density matrix afer Y}, then
the geometric discord is obtained as:

\begin{equation}\label{DGy}
    DG_{y}= \frac{1}{2}\mu_{2}^{2}(1-4\eta)^{2}.
\end{equation}
As we seen in Fig.8 in $ 0 <\theta <\frac{\pi}{2} $ by
increase of speed of observer geometric discord death is happen slower.
In the other words geometric discord for slower observers is more faster.
On the other hand in $ \frac{\pi}{2} <\theta <\pi $  by increase of speed
of observers geometric discord leads to death by more gradient. It means that
geometric discord death in this interval is occur later for higher speed observers.
According to this figure for every velocity of observers unless $ \theta=\frac{\pi}{2}$
geometric discord tend to death asymptotically.

\section{Time evolution of quantum discord of two relativistic particle
 state transmitted under Pauli channels}
\subsection{Pauli channel $ \sigma_{x}, \sigma_{z} $ }

If two relativistic particles transmitted through Pauli channel $ \sigma_{x} $ and $ \sigma_{z}$
then the time evolution of it is as Eq.(\ref{density matrix after sigma x}),
 Eq.(\ref{ro z}), respectively. There are x-states.
We choose the algorithm of Ref.\cite{Amazaher} to compute the quantum discord.
Therefore the eigenvalues of Eq.(\ref{density matrix after sigma x}), Eq.(\ref{ro z}) are:
$$
\varepsilon_{1}=\xi(1-\mu_{1(3)})
$$
$$
\varepsilon_{2}=\xi(1+\mu_{1(3)})
$$
$$
\varepsilon_{3}=\eta(1-\mu_{1(3)})
$$
\begin{equation}\label{eigenx}
    \varepsilon_{4}=\eta(1+\mu_{1(3)}).
\end{equation}
We can obtain the optimal conditional Von Neumann entropy as:
\begin{equation}\label{SCx}
    SC = -\frac{1+\phi}{2}\log_{2}\left(\frac{1+\phi}{2}\right) -\frac{1-\phi}{2}\log_{2}\left(\frac{1-\phi}{2}\right)
\end{equation}
where $ \phi= \max\left[1-4\eta, \mu_{1(3)},  \mu_{1(3)}(1-4\eta)\right] $.
Since the the Von Neumann entropy for the two subsystems of this state are as:
$ S(\rho_{x} ^{A})=S(\rho_{x} ^{B})=1 $ .
Therefore the quantum mutual information as a measure of total correlation is as:
\begin{equation}\label{Ix}
    I(\rho_{x})= 2+\sum_{i=1}^{4}\varepsilon_{i}\log_{2}\varepsilon_{i}
\end{equation}
where $ \varepsilon_{i} $'s are the eigenvalues in Eq.(\ref{eigenx}).
So the quantum discord of two relativistic particles after transmit trough
Pauli channels $ \sigma_{x},\sigma_{z} $ is given as:
$$
D(\rho_{x})= -1+\frac{\nu}{\ln16}\left(\ln\left[(8\nu)^{4}\xi^{3}\eta\right]+(8\xi-3)\ln\frac{\xi}{\eta}\right)
$$
\begin{equation}
+\frac{\vartheta}{\ln16}\left(\ln\left[(8\vartheta)^{4}\xi^{3}\eta\right]+(8\xi-3)\ln\frac{\xi}{\eta}\right)+SC
\end{equation}
where $ \nu=\sqrt{\mu_{1(3)}}\cosh(\gamma_{1(3)}t) $ and $ \vartheta=\sqrt{\mu_{1(3)}}\sinh(\gamma_{1(3)}t) $.
As we see in Fig.9 quantum discord is invariant for different
velocity of observers at first then it is decreased until it tend to DSD. It is interesting
that this reduction is more in $ 0 <\theta <\frac{\pi}{2} $ for lower speed observers but in
$ \frac{\pi}{2} <\theta <\pi $ the quantum discord reduction is more for higher speed observers.

\subsection{Pauli channel $ \sigma_{y}$ }

If two relativistic particles transmitted through Pauli channel $ \sigma_{y} $
then the time evolution of it is as Eq.(\ref{density matrix afer Y}) which is a x-state.
As in the previous section we can obtain the eigenvalues for
Eq.(\ref{density matrix afer Y}) as:
$$
    \epsilon_{1}=\epsilon_{2}=0
    $$
    $$
    \epsilon_{3}=\frac{1}{2}\left(-\mu_{2}\left(4\xi-1\right)+1\right)
    $$
    \begin{equation}\label{eigeny}
        \epsilon_{4}=\frac{1}{2}\left(\mu_{2}\left(4\xi-1\right)+1\right).
    \end{equation}
On one hand the Von Neumann entropy for marginal subsystems are given as:
$$
S(\rho_{y}^{A})=S(\rho_{y}^{B})=1
$$
Therefore the quantum mutual information for this system is as:
\begin{equation}\label{Iy}
    I_{y}=\frac{2}{\ln16}\left[\lambda_{2}\ln\left(\frac{1+\lambda_{2}}{1-\lambda_{2}}\right)+\ln4\left(1+\lambda_{2}\right)\left(1-\lambda_{2}\right)\right].
\end{equation}
On the other hand the classical correlation is 1.
Therefore we can obtain the quantum discord as:
\begin{equation}\label{QDy}
    QD(\rho_{y})=\frac{2}{\ln16}\left[\lambda_{2}\ln\left(\frac{1+\lambda_{2}}
    {1-\lambda_{2}}\right)+\ln\left(1+\lambda_{2}\right)\left(1-\lambda_{2}\right)\right].
\end{equation}
As e seen in Fig.10 in the interval $ 0<\theta<\frac{\pi}{2} $ the
quantum discord gradient reduction is more for lower speed observers.
In the other words lower speed observers tend to quantum discord death with more gradient.
On the other hand in the interval  $ \frac{\pi}{2}<\theta<\pi $ by increase of
speed of observer the quantum discord gradient reduction become more and more.
In the other words the higher speed observers tend to quantum discord death with more gradient in this interval.
\section{Conclusions}
In this paper we have investigated the evolutions of quantum entanglement and geometric discord
and quantum discord dynamics of two relativistic particles under noisy Pauli channels. we have found out
these quantum correlations dependence on speed of observer in Pauli channels.
We have shown that under each of Pauli channels $ \sigma_{x}, \sigma_{y}, \sigma_{z} $
quantum entanglement  and geometric discord
and quantum discord tend to death. This behavior occur for the lower speed observers in
$ 0 < \theta< \frac{\pi}{2} $ and in $ \frac{\pi}{2} < \theta< \pi $ these are vice versa.
The quantum entanglement sudden death(ESD) occur under Pauli channels $ \sigma_{x}, \sigma_{z} $.

\bibliographystyle{apsrev}
\newpage
\begin{figure}[htb]
\begin{center}
\includegraphics{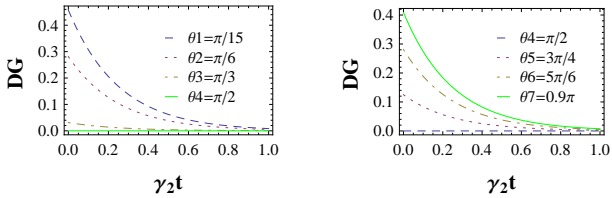}
      \vspace{6cm}
\caption{Geometric discord(DG) for $ 0 <\theta <\frac{\pi}{2}$ (left panel)
and geometric discord for $ \frac{\pi}{2}<\theta <\pi $ (right panel) measures
versus $\gamma_{2} t $  for different values of $ \theta $ under Pauli noisy channel
$ \sigma_{y} $ is sketched in this figure. As we seen geometric discord tend to death
for lower speed observers asymptotically more sudden in $ 0 <\theta <\frac{\pi}{2}$
 but for $ \frac{\pi}{2}<\theta <\pi $ is vice versa.}
\label{fig8}
\end{center}
\end{figure}
\begin{figure}[htb]
\begin{center}
\includegraphics{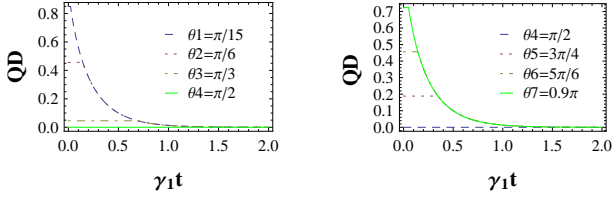}
      \vspace{6cm}
\caption{Quantum discord(QD) for $ 0 <\theta <\frac{\pi}{2}$ (left panel)
and quantum discord for $ \frac{\pi}{2}<\theta <\pi $ (right panel) measures
versus $\gamma_{1} t $  for different values of $ \theta $ under Pauli noisy channel
$ \sigma_{x} $ and $ \sigma_{z} $ which is same is sketched in this figure.
As we seen at first quantum discord is invariant
then quantum discord tend to death accordance together for all observers
. This reduction of quantum discord is more for lower speed observers
in $ 0 <\theta <\frac{\pi}{2}$ but for $ \frac{\pi}{2}<\theta <\pi $ is vice versa.}
\label{fig9}
\end{center}
\end{figure}
\begin{figure}[htb]
\begin{center}
\includegraphics{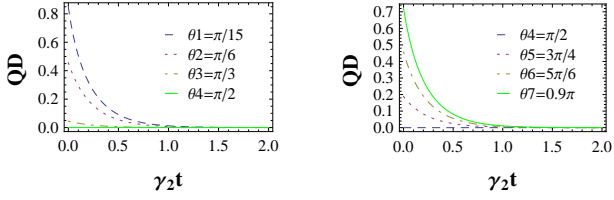}
      \vspace{6cm}
\caption{Quantum discord(QD) for $ 0 <\theta <\frac{\pi}{2}$ (left panel)
and quantum discord for $ \frac{\pi}{2}<\theta <\pi $ (right panel) measures
versus $\gamma_{2} t $  for different values of $ \theta $ under Pauli noisy channel
$ \sigma_{y} $ is sketched in this figure. As we seen in the interval $ 0<\theta<\frac{\pi}{2} $ quantum discord tend to
death asymptotically with more gradient for lower speed observers but in the interval  $ \frac{\pi}{2}<\theta<\pi $
is vice versa
.}
\label{fig10}
\end{center}
\end{figure}
\end{document}